\documentclass[aps,prl,twocolumn,superscriptaddress,floatfix,notitlepage]{revtex4-1}
\usepackage{amsmath}
\usepackage[T1]{fontenc}
\usepackage{graphicx}
\usepackage{hyperref}
\hypersetup{
    colorlinks=true,
    linkcolor=blue,
    filecolor=magenta,      
    urlcolor=blue,
}

\begin{document}

\title{Outflows from naked singularities, infall through the black hole horizon:
hydrodynamic simulations of accretion in the Reissner-Nordstr\"om space-time}

\author{W\l{}odek Klu\'zniak}
\email{wlodek@camk.edu.pl}
\author{Tomasz Krajewski}
\email{tkrajewski@camk.edu.pl}
\affiliation{Nicolaus Copernicus Astronomical Center of the  Polish Academy of Sciences,\\
Bartycka 18, 00-716 Warsaw, Poland}

\begin{abstract}
We performed the first simulations of accretion onto the compact objects in the Reissner-Nordstr\"{o}m (RN) spacetime. The results obtained in general relativity are representative of those for spherically symmetric naked singularities and black holes in a number of modified gravity theories. A possible application of these calculations is to the active galactic nuclei (AGNs) with their powerful jets. It is now possible to compare the results of such simulations with the accreting supermassive objects in our own Milky Way and the nearest spiral galaxy: observations of the core regions of galactic nuclei (Sgr A* and M87) performed with unprecedented resolution by the Event Horizon Telescope (EHT) collaboration allow fairly direct tests of the spacetime-metric of the central compact object. In this context we present general-relativistic hydrodynamical simulation results of accretion from an orbiting accretion torus (with a cusp) onto a RN black hole and a RN naked singularity. The results could not be more different for the two cases. For a black hole, just as in the familiar Kerr/Schwarzschild 
case, matter overflowing the cusp plunges into the black hole horizon. For the naked singularity, the accreting matter forms an inner structure of toroidal topology and leaves the system via powerful outflows. It is an open question whether this inner structure can give rise to an image quantitatively similar to the ones reported by EHT for M87 and Sgr~A*.

\end{abstract}

\maketitle
\section{Introduction} \label{sec:intro}
In this \emph{Letter} we investigate the difference between accretion onto a black hole and a spherically symmetric naked singularity. To this end we perform hydrodynamical simulations in the Reissner-Nordstr\"{o}m (RN) metric of an orbiting torus filled with perfect fluid, which overflows through a cusp in the self-intersecting equipotential surface, just as in the classic Kerr black-hole accretion case \citep{Abramowicz:1978}. The simulations are carried out with the general-relativistic code \texttt{Koral} \cite{Sadowski:2012}, which has been adapted \cite{Krajewski:2025} to the RN metric especially for this purpose.

The RN naked singularity is taken here to represent a wider class of horizonless objects in modified theories of gravity (in addition to singularities, these could also correspond to the so called regular black holes, e.g., \citep{Bardeen1968,Hayward:2006}). In fact, several of these theories yield a metric identical to the RN one, albeit with a different interpretation of the "charge" parameter $Q$; these include \footnote{These and other spacetime metrics are reviewed in ref. \citep{Vagnozzi2023}.} the BBMB theory \cite{Bocharova:1970, Bekenstein:1974sf}, Horndeski gravity \cite{Babichev:2017guv}, Randall-Sundrum II braneworld "black hole solution" \cite{Aliev:2005bi}, and Moffat's modified gravity \cite{Moffat:2014aja}.

Research into the astrophysical properties of putative naked singularities, and other alternatives \cite{Vagnozzi2023} to the Kerr spacetime, is topical in view of  recent advances in technology which allow fairly direct tests (under certain assumptions) of the spacetime metric in the center of our Galaxy, the Milky Way, and the nearby galaxy M87 \citep{EHT5,EHT2022}. Our study is also motivated in part by the lack of direct evidence (so far) as to the nature of the compact object in the powerful active galactic nuclei (AGNs) of distant galaxies. 
Galaxies contain supermassive objects at their centers. It is not yet clear how and when supermassive black holes may have been formed \citep{Inayoshi2020ARAA}. 

While the enormous luminosity of AGNs is understood in terms of accretion onto a supermassive black hole, and their highly relativistic jets (which may be powered by the rotation of a spinning black hole) are thought to be driven by magnetohydrodynamic effects \cite{Blandford:2018iot,Davis:2020wea},
the simple alternative of accretion flow onto a horizonless compact object has apparently not yet been considered.
Our surprising finding is that in contrast to black holes, powerful outflows are already a natural outcome of purely hydrodynamic infall into the vicinity of a non-rotating naked singularity.

A fundamental feature of the black hole is the presence of the event horizon, which cannot support any material structure abutting it. A naked singularity has no horizon, and in many theories of gravity, including the RN solution in general relativity (GR), it has a gravitationally repulsive core \citep{Pugliese:2010ps,VieiraSK:2014,ruchiphd} that can support structures (fluid bodies) in hydrostatic equilibrium \cite{VieiraK:2023cvn, MishraK:2023bhns.work..151M, MishraKK:2024}. 
As we show, this fundamental difference leads to a striking difference in the outcome of hydrodynamic accretion onto these two classes of compact objects. The black hole absorbs the accreting fluid, while the naked singularity does not. The naked singularity creates strong outflows by redirecting the infalling fluid, while the black hole does not.

\section{Numerical setup and the spacetime}
In this work we are interested in objects of astrophysical sizes, primarily stellar-mass objects and the supermassive compact objects in centers of galaxies. In this regime radii of the event horizon or the zero-gravity sphere will be of macroscopic scale, allowing us to use the fluid approximation for motion of matter in the considered systems. Our study is performed in GR, using the well understood formalism of relativistic hydrodynamics. We neglect the gravitational back-reaction coming from the fluid, assuming that the mass of the fluid torus is much smaller than the mass of the central object; otherwise, the fluid density in our non-radiative simulations is scale-free. 

With its radiative capabilities, the general relativistic magnetohydrodynamics code \texttt{Koral} \cite{Sadowski:2012}  has been successfully used to simulate accretion onto black holes \cite{Sadowski:2013gua,Lancova:2019ldt} and neutron stars \cite{Abarca:2021tzo}. In our simulations, we use an extension of the code to non-Kerr metrics, the \texttt{Koral+} code \cite{Krajewski:2025}. 
With this improved numerical tool, we investigated the motion of electrically neutral fluid in the RN metric background.
Here, we only report on non-radiative hydrodynamics runs.

The RN metric \cite{Reissner:1916,Nordstrom:1918} is a~vacuum solution of Einstein-Maxwell set of equations. In spherical coordinates $(t,r,\theta,\phi)$, for charge $Q$ \footnote{In fact the charge $Q$ does not need to be electromagnetic charge and may be the charge of any (hypothetical) Abelian gauge symmetry.} and mass $M$, the metric tensor $g$ has the form 
\begin{equation}
g =-f(r) dt^2+\frac{1}{f(r)}dr^2+r^2(d\theta^2+\sin^2\theta^2d\phi^2),
\label{eq:Reissner_Nordstrom_metric}
\end{equation}
with
\begin{equation}
-g_{tt}=f(r)=1-2M/r+{Q}^2/r^2, 
\label{eq:Reissner_Nordstrom_gtt}
\end{equation}
where $t$ is the time measured by a stationary clock at infinity. 
We use units such that $G=c=1$.

The radial position, $r=r_\mathrm{h}$, of the event horizon, if present, is defined by $g_{tt}(r_\mathrm{h})=0$. 
For the RN black hole there are two roots to this equation, only the larger one, at  $r_\mathrm{h}= M + \sqrt{M^2 - Q^2}$, is relevant to the outside observer. The radius of this outer horizon  varies monotonically with $Q^2$ from the Schwarzschild value of $r_\mathrm{h}= 2 M$ for the uncharged black hole ($Q=0$), to $r_\mathrm{h}= M$ for the extremal one ($Q^2=M^2$). For $Q^2 > M^2$ the metric function is always strictly positive and the spacetime does not have an event horizon at all; this case gives the naked singularity solutions. 

In the naked singularity case, there is a "zero-gravity" sphere of radius $r_0 = Q^2/M>M$ on which the acceleration of a test particle vanishes \cite{Pugliese:2010ps}.  For $r<r_0$,  gravity is effectively repulsive. This feature of the naked singularity allows for the presence of a levitating spherical (shell-like) atmosphere of zero-angular-momentum fluid \cite{VieiraK:2023cvn}. For fluid with non-zero angular momentum,  the repulsive nature of the gravitational field within the zero-gravity sphere of the naked singularity strongly influences the shape of toroidal figures of equilibrium \cite{MishraK:2023bhns.work..151M, MishraKK:2024}.

As there is no physical dissipative mechanism in our non-magnetized, perfect fluid simulations, for the initial condition we chose a situation allowing inflow of fluid with no loss of angular momentum or energy. Specifically, we used a fluid configuration corresponding to a rotating figure of equilibrium with a cusp \citep{MishraKK:2024}, which is formed by a self-intersection of the equipotential surface delimiting the fluid body. The fluid can overflow from the torus through the cusp if the (finite) outer volume enclosed by the self-intersecting equipotential surface is filled to the brim. The shape of the equilibrium torus  \cite{Abramowicz:1978} is determined  by the distribution of the  $l:= u_\phi / u_t$ parameter \footnote{The parameter $l$ is the specific angular momentum of a test particle.} in its interior, where the four-velocity of the fluid in the torus is $(u^t,0, 0, u^\phi)$ . In our simulations, the barotropic fluid was taken to initially have uniform distribution of $l$ equal to $l_0$. For the RN black hole, values of $l_0$ can always be found for which such a self-intersection occurs; for the RN naked singularity this is only possible for values of $Q/M$ rather close to unity \cite{MishraKK:2024}. Accordingly, we performed two simulations in the RN metric, one for $Q=1.02M$ (a naked singularity allowing a torus with a cusp), and the other for $Q=0.99M$ (a black hole). The reason we chose a nearly extremal black hole was to have very similar values of $Q/M$ for the two simulations. Outside some radius, say $r=2 M$, the value of $f(r)$ differs by no more than a few percent between the two cases. Thus, the initial configuration of the fluid is nearly identical in the two runs. It is only during the run that the infalling fluid can feel the dramatic difference in the geometry of the spacetime between the vicinity of the horizon and that of the zero-gravity sphere.

For the black hole we used $l_0=3.28$, the cusp  of the initial torus is then at $r=2.8 M$, its center (circle of maximum pressure) at $r=6.51M$. For the naked singularity with $l_0=3.22$ the cusp is at $r=2.65 M$, the center of the torus at $r=6.20M$. We used the Kerr-Schild coordinates, and outflow boundary conditions on the domain boundary, except for reflective conditions on the spherical inner boundary in the naked singularity case. The ideal gas equation of state was used in the runs, $p=(\Gamma-1)\varepsilon$, with $\Gamma =4/3$, where $p$ is the pressure and $\varepsilon$ is the specific internal energy. The runs are purely hydrodynamic, no magnetic fields or radiation are included. More numerical details can be found in supplemental materials.

We report runs of length $5 \times 10^4\, t_\mathrm{g}$. All times $t$ correspond to the time elapsed after initialization, and are given in units of the gravitational time $t_\mathrm{g}=M$.

\section{Accretion onto a black hole}

\begin{figure}[!p]
\centering 
\includegraphics[width=\columnwidth]{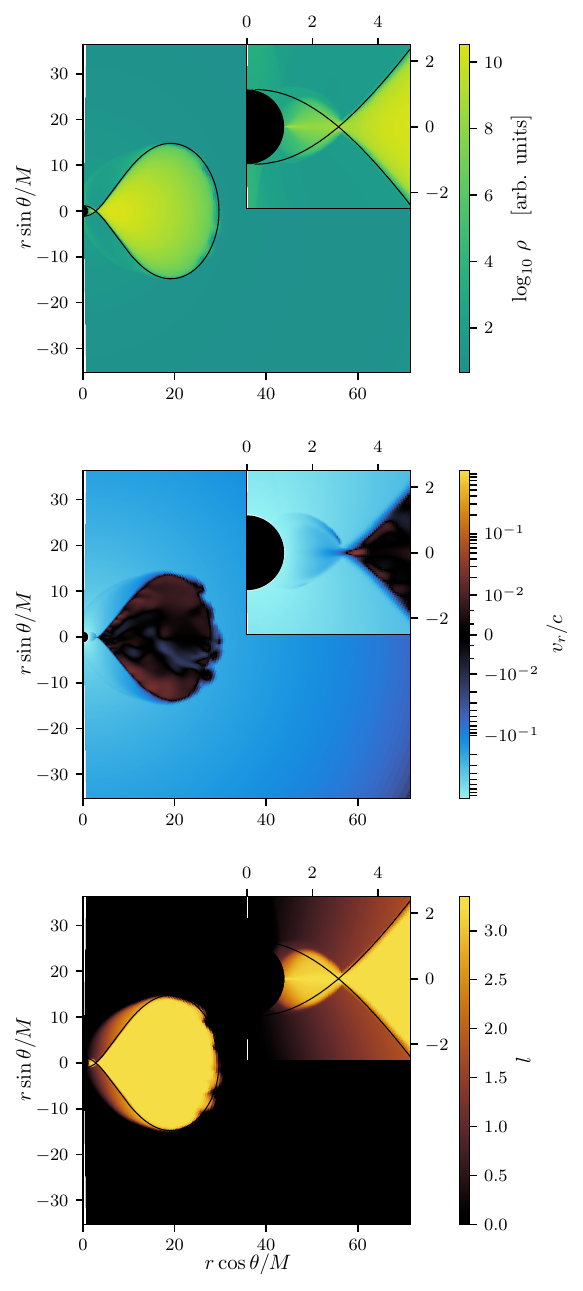}
\caption{Steady-state accretion onto a $Q/M=0.99$ Reissner-Nordström black hole. A snapshot of the simulation at time $t=10^4\, t_\mathrm{g}$ is shown. {\sl Top panel:} $\log_{10} \rho$, the logarithm of the rest mass density. {\sl Middle panel:} radial component of the fluid velocity $v_r=-u^r/u_t$ in units of the speed of light, $c$. {\sl Bottom panel:} the value of $l=u_\phi/u_t$.  Clearly, no fluid is ejected from the system. The black circular disc (clearly seen in the insets) corresponds to the (outer) event horizon. The black solid line in the top and bottom panels indicates the self-intersecting equipotential surface for $l_0=3.28$.}
\label{fig:q=0.99_rcusp=2.8_snapshot}
\end{figure}

The black hole simulation is used as a benchmark for the naked singularity one. The initial torus is very close to hydrostatic equilibrium, with zero poloidal velocity. However, close to the surface some fluctuations may occur, and at any rate some of the gas crosses the cusp. After some time, of the order of $t=10^3\, t_\mathrm{g}$ conditions  stabilize into a nearly steady-state flow. A snapshot of this state is presented in Fig.~\ref{fig:q=0.99_rcusp=2.8_snapshot}.

\begin{figure}[!p]
\centering
\includegraphics[width=\columnwidth]{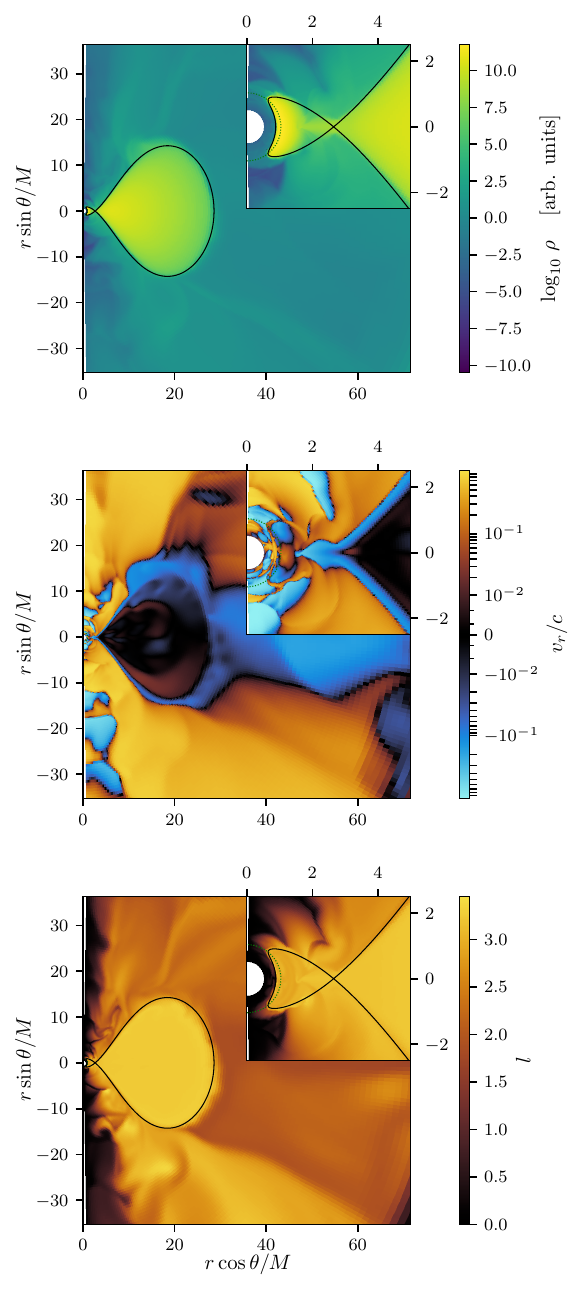}
\caption{Steady flow pattern in accretion onto a $Q/M=1.02$ Reissner-Nordström naked singularity placed at the origin of the coordinate system. A snapshot at time $t=5 \times 10^4\, t_\mathrm{g}$ is shown. 
The panels show the same quantities as in Fig.~\ref{fig:q=0.99_rcusp=2.8_snapshot}.
 Note the mildly relativistic outflow, generally away from the equatorial plane. 
The white disc is outside the computational domain. The black solid line in the top and bottom panel indicates the self-intersecting equipotential surface for $l_0=3.22$. The green dotted line in the insets corresponds to the zero-gravity sphere.}
\label{fig:q=1.02_rcusp=2.65_steady_state}
\end{figure}

The outcome of the hydrodynamical solution in the black hole case is very simple, and its interpretation is straightforward. The fluid overflows the torus through the cusp. The torus slightly expands, perhaps also as a result of numerical imperfections of the model, and some gas crosses the equipotential surface in its part facing the axis of symmetry (outside the cusp). The flow is fairly laminar. All the gas outside the torus, including the numerically necessary low-density background, has a negative radial velocity, i.e. it flows towards the black hole. Eventually, it crosses the event horizon and is never seen again. 

One implication of our black hole hydrodynamic run is that in real black holes, jets and outflows \emph{require} magnetic fields, a spinning black hole, and/or radiative effects. Without them, as can be seen in Fig.~\ref{fig:q=0.99_rcusp=2.8_snapshot}, the black hole simply consumes all the fluid in the accretion flow.

\section{Flow about a naked singularity
\label{sec:naked_singularity}}

In our simulation of outflow from a torus around the $Q/M=1.02$ RN naked singularity,  the flow pattern also stabilizes at $t\sim10^3\, t_\mathrm{g}$. Snapshots from the simulation at time $t=5 \times 10^4\, t_\mathrm{g}$ are presented in Fig.~\ref{fig:q=1.02_rcusp=2.65_steady_state}. 

As before, the fluid streams through the cusp towards the central compact object. Again, some fluid may also possibly cross the equipotential surface outside the cusp. What is very apparent in the middle panel of Fig.~\ref{fig:q=1.02_rcusp=2.65_steady_state} is that the flow pattern is dominated by outflows (positive values of $v_r$). However, there is an inflowing stream (negative values of $v_r$) of matter sliding along the surface of the torus.
The fluid does not accrete onto the naked singularity. Owing to the repulsive nature of the gravitational field in the close neighbourhood of the singularity, a part of the fluid accumulates near the zero-gravity sphere (but not at high latitudes), the remainder is deflected and outflows from the system at mildly relativistic velocities (up to about $0.8 c$ in the funnel around the axis of symmetry). However, the bottom panel clearly shows that the immediate vicinity of the $z$ axis is free of any matter coming from the torus (it is filled, instead, with the numerically necessary low density background).

The matter accumulating outside the naked singularity forms another torus (of crescent-like cross-section). This inner torus
is clearly visible in the top panel of Fig.~\ref{fig:q=1.02_rcusp=2.65_steady_state}. By inspection of the bottom panel of Fig.~\ref{fig:q=1.02_rcusp=2.65_steady_state}  one may deduce that the fluid in this inner torus has a nearly uniform value of $l$.

The accretion stream formed by the plasma overflowing through the cusp of the outer torus initially preserves its original value $l_0$. However, as is visible in the inset, the material in the inner torus has a somewhat lower value of $l$ than in the outer torus.

A characteristic feature of flow about the naked singularity is the strong outflow of the fluid that has overflowed from the orbiting initial torus. Not only is there a significant outflow of material, as visible in the middle panel of Fig.~\ref{fig:q=1.02_rcusp=2.65_steady_state}, but the accumulating and swirling fluid seems to influence the accretion stream leaving the cusp, breaking the equatorial plane-reflection symmetry of flow to a much higher degree than in the black hole case. Indeed, as is obvious from inspection of Fig.~\ref{fig:q=0.99_rcusp=2.8_snapshot}, all the matter leaving the orbiting torus in our black hole simulations accretes onto the event horizon and can no longer influence the dynamics of accretion. 

The process of accretion onto naked singularities described by the RN metric proceeds in a qualitatively similar fashion for other values of parameter $|Q| / M$, as we discuss in detail for $1 < |Q| / M < \sqrt{5} / 2$  in a full-length paper \citep{Krajewski:2024}. Simulations with $|Q| /M > \sqrt{5} / 2$ also present similar behaviour, however, they have to take into account dissipative processes in the fluid that transport angular momentum, since no equilibrium torus for such values of $|Q| / M$ can develop a cusp from which the matter could accrete without loss of angular momentum.

\section{Summary and discussion}
\label{sec:summary}
We investigated accretion of an electrically neutral fluid onto spherically symmetric black holes and naked singularities. We found that the fundamental difference in the existence, or not, of the event horizon causes equally fundamental differences in the pattern of accretion. Qualitatively, the flow is completely different in the two cases. Our findings may be relavant to observations of AGNs.

We used the GR hydrodynamical numerical code \texttt{Koral} \cite{Sadowski:2012} to solve for the dynamics of a perfect fluid accreting onto the investigated compact objects. For simplicity, the Reissner-Nordström background metric was assumed, so that we could use the familiar equations of GR hydrodynamics, and also because with a proper choice of the value of $Q$ parameter the RN metric can model both (charged) black holes and naked singularities.

In our naked singularity simulation, a toroidal structure with a crescent-like  cross-section is formed near the zero-gravity sphere. At the same time, powerful outflows are present in our purely hydrodynamic simulation already, a fraction of the infalling fluid having turned around leaves the system with mildly relativistic velocities. We infer, that RN-like naked singularities are conducive to forming jets and outflows. We suggest that the ray-traced image of the inner toroidal structure be compared with the direct the Event Horizon Telescope observations of Sgr A*.

In contrast, the black hole in our simulation accretes the inflowing matter, all of it being absorbed by the event horizon. Thus, any outflows or jets require the presence of radiation, magnetic fields or of a rotating black hole (all of which are absent in the presented simulations). 

We end by noting that the propagation of radiation in the vicinity of a naked singularity may be significantly influenced by the matter moving about it. This may affect studies of general naked singularities (e.g. \cite{daSilva:2023jxa,Viththani:2024fod}), and even possibly of Kerr naked singularities \cite{Tavlayan:2023vbv,Mummery:2024znv}, especially at the critical spin value of $a_*\sim1$. In particular, images of the naked singularity may be strongly modified by the opaque material in the powerful outflows and the crescent-like inner torus that we report here. The evolution of the naked singularity may also not be as expected. If we are not mistaken, most previous studies of accretion onto naked singularities assumed that the fluid leaving the accretion disk (through the marginally stable orbit, when present) will accrete onto the naked singularity, changing its mass and angular momentum (or other parameters). At least for the spherically symmetric naked singularities we are considering here, this does not seem feasible.

\begin{acknowledgments}
Research funded in part by the Polish NCN grants No. 2019/33/B/ST9/01564 and 2019/35/O/ST9/03965. High-performance computations were performed on the CHUCK cluster at CAMK, Warsaw. The authors thank Debora Lan\v{c}ov\'a and Miljenko \v{C}emelji\'{c} for stimulating discussions. We thank Ruchi Mishra and Angelos Karakonstantakis for testing the numerical code in early stages of this work. 
\end{acknowledgments}

\bibliography{references}

\end{document}


\title{Outflows from naked singularities, infall through the black hole horizon:
hydrodynamic simulations of accretion in the Reissner-Nordstr\"om space-time\\
\emph{Supplemental Material}}

\author{W\l{}odek Klu\'zniak}
\email{wlodek@camk.edu.pl}

\author{Tomasz Krajewski}
\email{tkrajewski@camk.edu.pl}
\affiliation{Nicolaus Copernicus Astronomical Center of the  Polish Academy of Sciences,\\
Bartycka 18, 00-716 Warsaw, Poland}

\maketitle

\section{Numerical methods used in \texttt{Koral+} code\label{sec:numerical_code}}
Results presented in the \emph{Letter} were obtained using the \texttt{Koral+} code, which is an~extension of the well-known \texttt{Koral} \cite{Sadowski:2012, Sadowski:2013gua}. The version of the code that we used is an intermediate step in the ongoing, long-term project of modernizing the \texttt{Koral} legacy code. Up to now, the metric-dependent part of the code was refactorized and the new part simplifying implementation of new systems of coordinates exploiting symbolic computations software was added.

Computations in gravitational background of the black holes required horizon penetrating coordinates system like spherical Kerr-Shild coordinates $(\tau, r, \theta, \phi)$ where
\begin{equation}
    dt = d\tau - \frac{2 M r}{r^2 - 2M r} dr
\end{equation}
outside horizon. In our simulations we used the modified spherical Kerr-Shild coordinates which has radial coordinate $s$ logarithmically stretched, i.e. $s = \log{r}$ with respect to the standard Kerr-Shild coordinates $(\tau, r, \theta, \phi)$.
We used the same coordinate system for our naked singularity simulations.

In our simulations we assume exact axial symmetry which allows us to reduce the problem to two spatial dimensions. The computational domain for simulations of accretion onto the black hole is $\log(0.90) < s < \log(80)$ and $0.01 \pi < \theta < 0.99 \pi$, thus a part of computational domain is below event horizon. For simulations of naked singularity our choice is $ \log(0.52) < s < \log(80)$ and $0.01 \pi < \theta < 0.99 \pi$. Boundaries of computational domains have four parts. Boundary conditions for black hole simulations at inner boundary, i.e. $s = \log(0.90)$ were outflowing ones, while for the naked singularity simulations boundary at $s = \log(0.52)$ we have chosen reflective boundary conditions. This choice is motivated by the observation that test particles are repelled from the vicinity of the naked singularity. However, the accreting fluid hardly ever approaches this boundary. Boundary conditions at the outer boundary, i.e. $s = \log(80)$ and azimuthal boundaries, i.e. $\theta = 0.01 \pi, 0.99 \pi$ are the same for both setups. We allow outflow of the fluid through the outer boundary, while reflective boundary conditions are assumed at the azimuthal ones to mimic the true topology of the problem. The computational grid is formed by the equal division of $s$ coordinate interval into $256$ subintervals and $\theta$ coordinate interval into $256$ forming $256^2$ finite volumes.

The grid-based numerical schemes used by \texttt{Koral}(\texttt{+}) code are not able to simulate vacuum. Hence, one is forced to add artificial atmosphere outside the tori in initial conditions of the simulations. The atmosphere is formed by the fluid at rest with mass density much smaller than the density in the torus.

For our simulations we used the ideal equation of state
\begin{equation}
    p = (\Gamma - 1) \varepsilon,
\end{equation}
where $\Gamma= 4/3$ and $\varepsilon$ is the internal energy density of the fluid which is defined through the enthalpy as $w = \rho + \varepsilon + p$ with $\rho$ being the mass density.

The remaining numerical methods follow from \texttt{Koral} code and have been already described in details in \cite{Sadowski:2012, Sadowski:2013gua}. We will only briefly present our numerical setup.

Equations of relativistic hydrodynamics can be derived from conservation 
\begin{equation}
    \nabla_\mu T^{\mu \nu} = 0 \label{eq:energy_momentum_conservation}
\end{equation}
of energy-momentum tensor of the perfect fluid
\begin{equation}
    T^{\mu \nu} = w u^\mu u ^\nu + p g^{\mu \nu}
\end{equation}
which in astrophysical context is usually supplemented with rest mass conservation
\begin{equation}
    \nabla_\mu \left( \rho u^\mu \right) = 0. \label{eq:rest_mass_conservation}
\end{equation}
\texttt{Koral}(\texttt{+}) solves Eqs. \eqref{eq:energy_momentum_conservation} and \eqref{eq:rest_mass_conservation} in the form
\begin{subequations}
    \label{eq:conservation_form}
    \begin{align}
        \partial_t T^{t \nu} + \partial_i T^{i \nu} = -\Gamma^{\nu}_{\phantom{\nu} \mu \lambda} T^{\mu \lambda} - \frac{1}{\sqrt{-g}} \partial_i \left(\sqrt{-g}\right) T^{i \nu},\\
        \partial_t \left( \rho u^t\right) + \partial_i \left( \rho u^i \right) = - \frac{1}{\sqrt{-g}} \partial_i \left(\sqrt{-g}\right) \left( \rho u^i \right)
    \end{align}
\end{subequations}
which is accessible to standard Godunov-type numerical schemes for conservation equations.

\texttt{Koral}(\texttt{+}) is based on the method of lines in which equations \eqref{eq:conservation_form} are first discretized in space and then integrated in time using time stepping approach. The spatial discretization starts with conversion of conserved quantities \footnote{Here, we only describe and use the hydrodynamical part of the code. However, \texttt{Koral+} as its ancestor \texttt{Koral} is able to solve also equations of magneto-hydrodynamics coupled to radiation, using the M1 closure scheme for the latter.}:
\begin{equation}
    U = [\rho u^t, T^t_{\phantom{t}t} + \rho u^t, T^t_{\phantom{t}i}, S u^t]
\end{equation}
to so-called primitive ones:
\begin{equation}
    P = [\rho, \varepsilon, u^i, S],
\end{equation}
where $S$ is the entropy of the fluid which satisfies the following conservation equation
\begin{equation}
    \nabla_\mu \left( S u^\mu \right) = 0. \label{eq:entropy_conservation}
\end{equation}
The set of equations \eqref{eq:energy_momentum_conservation} and \eqref{eq:rest_mass_conservation} supplemented with \eqref{eq:entropy_conservation} is over-specified and \texttt{Koral}(\texttt{+}) solves \eqref{eq:entropy_conservation} only to use evolved entropy in a backup procedure \cite{Sadowski:2013gua} of obtaining primitive variables from conserved ones if the default algorithm "$1D_W$" of \cite{Noble:2005gf} fails. In the opposite case when conversion based on $\rho u^t$, $T^t_{\phantom{t}t}$, $T^t_{\phantom{t}i}$ succeeds the entropy is calculated from $\rho$ and $p$ as
\begin{equation}
    S = \frac{\rho}{\Gamma - 1} \log{\left( \frac{p(\varepsilon)}{\rho^\Gamma} \right)}
\end{equation}
and evolved up to the end of the timestep.

After the conversion the obtained primitive variables are corrected up to assumed numerical floors. The most important for our simulations are minimal value of $\rho$ which we assume is $10^{-50}$ is simulation units. This value is usually set in the interior of zero-gravity sphere. Furthermore, the diluted plasma contained in zero-gravity sphere is easily heated up by the accreting material which is limited by the maximal $\varepsilon / \rho$ ratio equal to $100$. Motivated by the findings of \cite{Siegel:2017sav} we limited maximal relativistic $\gamma = u^t / \sqrt{-g^{tt}}$ to $10$. This choice influences the dynamics of the relaxation of initial conditions inside the zero-gravity sphere, but is not important for later process of accretion of material from torus. Certain finite volumes during performed simulations are prone to obtain negative internal energy density $\varepsilon$ during conversion step. We decided to fix such pathological cells by substituting problematic values by the averages from neighbouring cells of the computational mesh. Due to high resolution (especially around zero-gravity sphere where such situation takes place) this fix-up procedure does not spoil the physical results obtained in the simulations.

The conversion of conserved quantities to primitive ones is followed by the reconstruction of quantities located at faces of finite volumes from volume averages. This means that \texttt{Koral}(\texttt{+}) reconstructs the primitive quantities and not the conserved ones. In our simulations we used linear interpolation for reconstruction corrected by the van-Leer’s minmod limiter with the dissipation parameter $\theta_\text{minmod} = 1.5$ \cite{Kurganov:2000241}. After limiting the numerical floors are applied to reconstructed quantities.

Reconstructed primitive variables are used to calculate fluxes using approximate Riemann solvers. For this research project we chose Harten, Lax and van Leer (HLL) solver \cite{Harten:1983}.

Calculated fluxes and sources from right hand sides of \eqref{eq:conservation_form} are used to integrate conservation equations in time by explicit midpoint method which is of second order in time step length.

\bibliography{references}